# High aspect ratio arrays of Si nano-pillars using displacement Talbot lithography and gas-MacEtch

Zhitian Shi[a,b], Konstantins Jefimovs[a], Marco Stampanoni[a,b], Lucia Romano[a,b]

[a]Paul Scherrer Institut, 5232 Villigen PSI, Switzerland
[b]Institute for Biomedical Engineering, University and ETH Zürich, 8092 Zürich, Switzerland

**Abstract** (Max 250 words)

Structuring Si in arrays of vertical high aspect ratio pillars, ranging from nanoscale to macroscale feature dimensions, is essential for producing functional interfaces for many applications. Arrays of silicon 3D nanostructures are needed to realize photonic and phononic crystals, waveguides, metalenses, X-ray wavefront sensors, detectors, microstructures and arrays of Si pillars are used as bio-interfaces in neural activity recording, cell culture, microfluidics, sensing and on-chip manipulation. Here, we present a new strategy for realizing arrays of protruding sharp Si nanopillars using displacement Talbot lithography combined with metal-assisted chemical etching (MacEtch) in gas phase. With the double exposure of a linear grating mask in orthogonal orientations and the lift-off technique, we realized a catalyst pattern of holes in a Pt thin film with a period of 1 µm and hole diameter in the range of 100-250 nm. MacEtch in gas phase by using vapor HF and oxygen from air allows to etch arrays of protruding Si nanopillars 200 nm-thick and aspect ratio in the range of 200 (pillar height/width) with an etching rate up to 1 µm/min. With the advantage of no stiction, no ion beam damage of the Si substrate, nanometric resolution and high fidelity of pattern transfer the method is an easy-to-scale-up processing that can support the fabrication of Si pillars arrays for many valuable applications both at micro and nano-scale.

# Introduction

In the last two decades, nanofabrication of high-aspect ratio 3D structures such as nanowires has been largely demonstrated with both bottom-up (whiskers growth[1]) and top-down (etching[2]) methods, contributing to the spreading of their applications in many different fields. Nanosized vertical silicon structures with high aspect ratio are used nowadays as functional interfaces for many different purposes, covering applications with a variety of diverse properties. For energy applications, Si nanowires as anode in a lithium battery enhance the battery charge capacity and promote a longer battery life[3,4]; nanowire based solar cells have been explored to reduce optical loss, enhance optical absorption, and improve carrier extraction for high performance and low cost designs[5]; Si micropillars and conformal junctions around them passivate the surface as well as improving charge collection in betavoltaics[6]. In thermoelectric devices Si has superior figure of merits[7], silicon based pillared phononic crystals[8] and nanophononic metamaterials[9] can serve as low-cost thermoelectric material with exceptional performance. Optical devices take advantages of arrays of semiconductors 3D nanostructures such as diffractive gratings and waveguides, benefiting from tunable light emission/absorption in a broad range of the spectrum from IR to UV. To pave the way for their widespread use, metalenses[10] need high-throughput and low-cost manufacturing by nanoimprint lithography[11], which is a replication process requiring a high aspect ratio master, typically produced by etching silicon in nanopillars arrays. Plasmonic enhancement has outlined the possibility to locally enhance absorption in selected wavelength range for better detectors, photovoltaics, and sensors. 3D arrays have been used as plasmonic nanostructures for surface enhanced Raman spectroscopy[12,13]. In addition, high aspect ratio pillar arrays are used in photodiodes[14], CMOS[15], X-ray wavefront sensors[16] and 3D neutron detectors[17]. Silicon nanowires can be used as platform for quantum computing experiments[18]. Nanostructures open opportunities for labeling and optical-based detection of biological species that offer advantages compared with conventional organic molecular dyes widely used today. The electronically switchable properties of semiconducting wires provide a sensing modality – direct and label-free electrical readout – that is exceptionally attractive for many applications, such as pH sensing, detection of proteins, DNA and virus[19]. Microstructures and arrays of Si pillars are used as bio-interfaces in neural activity recording[20], cell culture[21], microfluidics for DNA analysis to effectively increase the surface area of the channels[22], needles[23,24] and on-chip manipulation[25]. Microneedle technologies have the potential for expanding the capabilities of wearable health monitoring from physiology to biochemistry[26,27]. Nano-pillar arrays can enable the quantification of subnuclear abnormal features in tumor cells opening a new possibilities in characterizing malignant cells[28]. Materials patterned with high-aspect-ratio nanostructures have features on similar length scales to cellular components. Through careful design of their nanoscale structure, these systems act as biological metamaterials[24], eliciting unusual biological responses.

In such a broad range of applications, we are lacking a reliable and scalable technology platform not only to easily prototype and test the device properties in laboratory, but also to open the route towards device integration and fabrication. Ideally, we look for a cost-effective and easy-to-scale-up processing that can support the fabrication at both micro and nano-scale. Majority of the nanostructures are fabricated using electron beam lithography that is slow and expensive. Alternative methods have been developed to enable nanoscale fabrication faster and less expensive, such as nanoimprint lithography[29], that allows the realization of printed features even smaller than the initial master[30]. Displacement Talbot lithography (DTL) can quickly and reliably print periodic sub-micrometer arrays on a large scale below the resolution of a standard UV lithography[31-33]. Like steppers and interference lithography, DTL is essentially a contactless method with a large depth of focus, but with additional advantages, such as distortion-free control of the grating pitch and phase defined by the mask, compactness of the system and yet, with potential 3D capabilities of Talbot lithography.[34]

Etching is an important step in semiconductor device processing, conventional dry- and wet-etching methods have limited aspect ratio at nanoscale[35], metal assisted chemical etching (MacEtch) has been proposed[36,37] to produce silicon nanowire arrays with defined geometry and optical properties in a manufacturable fashion, demonstrating higher anti-reflectivity[38] and better absorption efficiency[37] in solar cells. To obtain Si nano-/microstructures with heights of over 1 μm, reactive ion etching (RIE) has typically been used to etch Si bulk films[5]. MacEtch provides a viable path for manufacturing of 3D functional semiconductor nanostructures with lateral resolutions down to 10 nm and extremely high aspect ratio, which is limited mainly by etching time[39]. While the lithographic methods led to sub–10-nm line resolution in planar devices, dry-etch cannot replicate such resolution out-of-plane due to (i) poor control over mask selectivity during etching, (ii) roughness produced via scalloping effects, and (iii) etch rate dependence on feature size. MacEtch allows to fabricate hierarchical features[40] with very sharp profile and has been successfully demonstrated in combination with electron beam[41,42], UV[39,43-45], flow-enabled selfassembly[46], tip-based[47], electrochemical nanoimprint[48] and chemisorption-assisted transfer printing[49] lithography. Our new method of MacEtch in gas phase (gas-MacEtch)[40], combines the extremely high aspect ratio capability of this technique with the use of gas reactants to reduce the nanostructures stiction due to capillary effects during drying after wet etching[50]. In this study, we demonstrate the combination of the wafer scale fast nano-patterning of DTL with gas-MacEtch to fabricate ordered arrays of protruding Si nanopillars with extremely high aspect ratio (> 200) and controlled feature size of about 200 nm. This method can be used to manufacture protruding high aspect ratio nanostructures in silicon substrates at multiscale[51] and applied, for example for the direct nano-fabrication in silicon of X-ray optics[41,44], bio-interfaces or nano/microprinting master stamps.

# Results and discussion

A schematic illustration of the combined process of DTL and gas-MacEtch to produce a periodic array of high aspect ratio Si vertical and protruding nanopillars in a <100> Si substrate is shown in Figure 1. First, a photoresist stack is spin-coated layer by layer on top of the Si substrate. The stuck includes a positive photoresist, a layer of Polymethylglutarimide (PMGI) to facilitate the lift-off process, and a bottom anti-reflective coating (BARC) that eliminates the stand wave effect during the exposure. Then, the photoresist is exposed through a mask with grating lines, the mask period is double of the desired period in the dots array. Because of interference, in DTL the photoresist is impressed with a pattern that has half of the period of the mask. The wafer is then rotated by 90 deg with respect to the mask for a second exposure. The double DTL exposure through a linear grating mask with period *2p* creates a periodic dots pattern into the photoresist stack with period *p*. During development, the exposed photoresist is dissolved and a periodic pattern of nanopillars with dose dependent pillar width is left (Fig.1.b). The pillar width is additionally reduced during BARC etch by an oxygen plasma. Finally, a Pt thin film (10-12 nm) is deposited and then lifted-off by dissolving the photoresist in acetone. As a result, a Pt film with regular hole arrays is left on the substrate surface (Fig.1.c), acting as catalyst during the subsequent gas-MacEtch process. The MacEtch reaction occurs in a gas-phase since the Pt patterned sample is suspended above a tank with a HF containing solution. Several attempts of MacEtch from a vapor source of HF have been reported with different catalysts, such as Au[52], Ag[53], TiN[54] and Pt[40]. The reaction gas is a mixture of HF with air, where HF is evaporated from the liquid solution and the oxygen is supplied by the air flow in the reaction chamber. The Pt film sinks into the Si substrate during the MacEtch reaction and an array of Si nanopillars is created (Fig.1.d). Since the process occurs at temperature higher than 40 °C, the liquid condensation on the Si nanostructures during the MacEtch process is prevented and the stiction or the deflection of the Si nanopillars due to liquid drying is minimized with respect to the typical wet-MacEtch process[39]. The cross-sectional size of the Si nanopillars corresponds to the size of photoresist pillars created by DTL, the height of the Si nanopillars is determined by the etching time in gas-MacEtch. The aspect ratio of the Si nanopillars is defined as the ratio between the height and the cross sectional width of the Si pillar. Arrays of vertical Si nanopillars with very high aspect ratio (>100) are easily manufactured with this protocol.

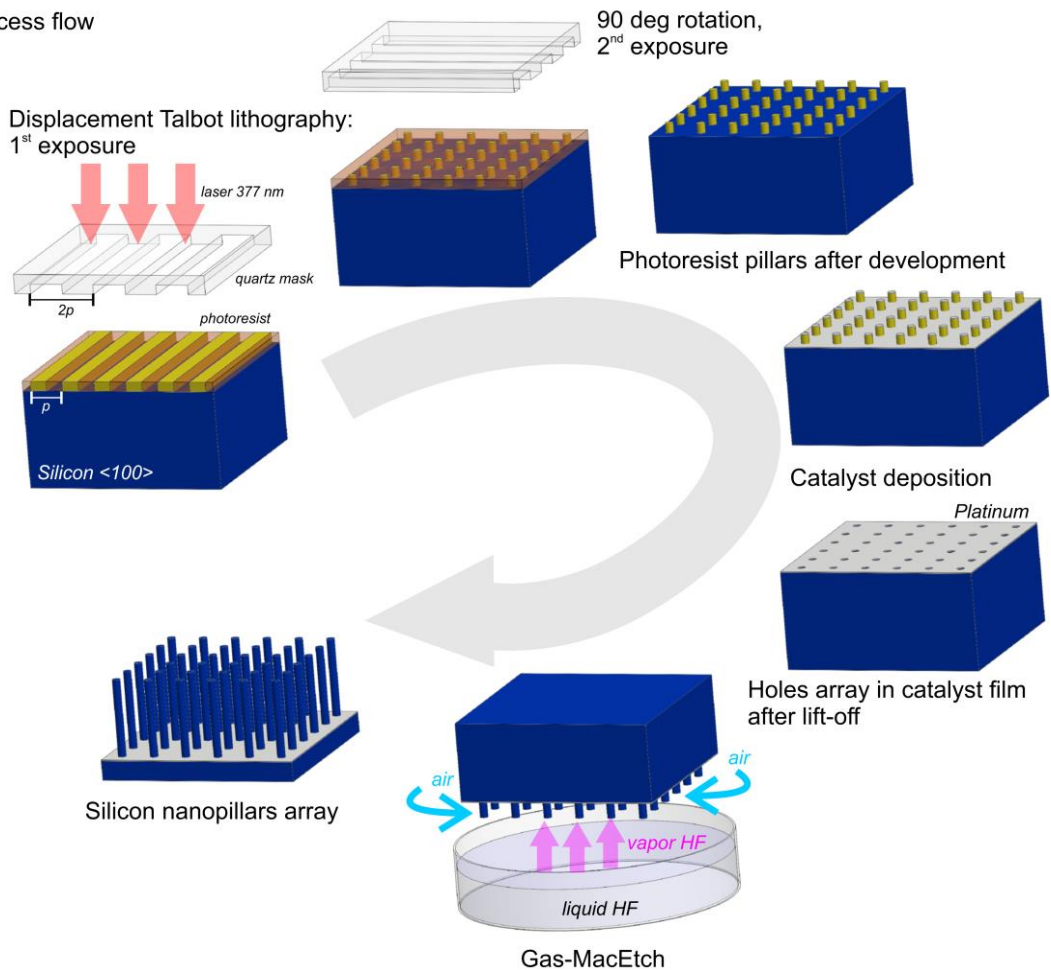

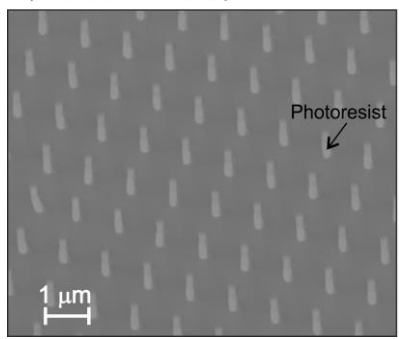 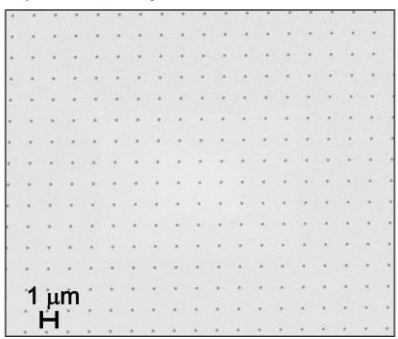 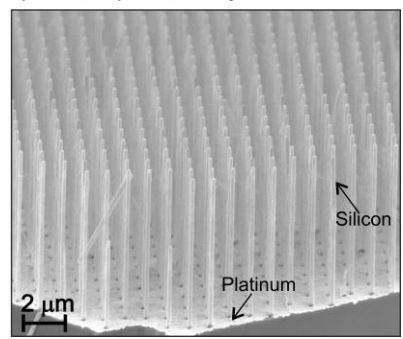

*Figure 1.* (a) Process flow of Si nano-pillar arrays fabrication in Si <100> substrates. Examples of different process steps, the double exposure DTL was used with a linear grating of pitch 2 μm (2*p*): (b) SEM in tilted view of photoresist nanopillar (width ~300 nm, height ~700 nm ) array with pitch 1 μm (*p*) after development; (c) plan view SEM of hole (hole diameter ~200 nm) arrays in Pt film after lift-off; (d) a typical bird's-eye-view SEM image of Si nano-pillar array with aspect ratio of 50 (height/width of the Si pillar) after gas-MacEtch.

**Displacement Talbot lithography with double exposure**

The dose distribution map of the double DTL process is simulated with Matlab. A fast Fourier transformation (FFT) method assuming Fresnel approximation is adopted for the simulation of the interference pattern generated by the DTL system. The essential parameters used for the simulation include: the wavelength of the laser light source ($\lambda$), the period of the mask ($p_{mask}$). The Talbot effect is observed when monochromatic coherent light is diffracting on a periodic structure (e.g. a grating) and forming so called Talbot carpet – self-images of the masks at a periodic distance (Talbot distance $z_T$) due to interference of light from different diffraction orders. The Talbot distance is calculated with the following equation:

$$z_T = \frac{\lambda}{1 - \sqrt{1 - \frac{\lambda^2}{p_{mask}^2}}} \quad (1)$$

The intensity along the laser propagation direction is integrated within one or several Talbot distances $z_T$ resulting in high-contrast image, which is independent of initial separation of mask and wafer[33]. The dose distribution map of the first DTL exposure is plotted in Figure 2. The laser source has a Gaussian like spectral distribution, hence a Gaussian filter $\sigma$ is applied to the dose distribution map.

.

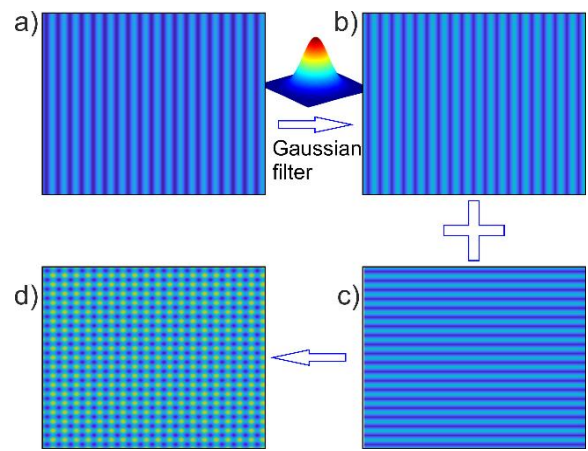

*Figure 2.* Simulated double exposure in DTL: (a) Dose distribution map with linear grating mask without considering laser Gaussian-like spectrum distribution, (b) actual dose distribution map after first DTL exposure, (c) a second DTL exposure but rotating 90°, the same Gaussian filtering is applied like Fig.2(b), (d) final dose distribution map after double DTL exposures.

As shown in Figure 2.b, the limited monochromaticity of the laser source smears the dose map. In linear patterns this effect produces only a marginal reduction of duty cycle, which can be easily moderated by adjusting the exposure time. However, the effect is more complicated in a double DTL exposure. After the first exposure is completed, we rotate the wafer by 90° and do a second DTL exposure. Due to a Gaussian blur, the second exposure (Fig. 2.c) on top of the first one (Fig. 2.b) produces a pattern that contains dots with rounded corners in a staggered arrangement (Fig. 2.d). Figure 3 reports the simulated evolution of the size and the shape of the dots by applying Gaussian filters with different $\sigma$ values and exposure time.

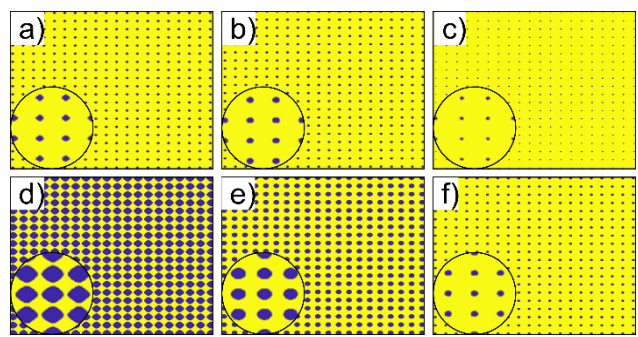

*Figure 3.* Dual DTL exposure results. Simulations: (a) – (c) different $\sigma$ values of the Gaussian filters ($\sigma$ = 1, 5, 10); (d) – (f) different exposure time (1×, 2×, 3× unit dose).

The $\sigma$ value is determined by several factors, including the properties of the laser light source and the photoresist that is used for the exposure. The $\sigma$ value is fixed once the process is established, therefore it cannot be used to control the pattern. Other than that, another important factor that has a direct influence on dots' size and shape is the overall light exposure, which is determined by the total exposure time. In Figure 3.d-f the size and the shape of the dots drastically change with increasing the dose. This is confirmed by the experimental observations.

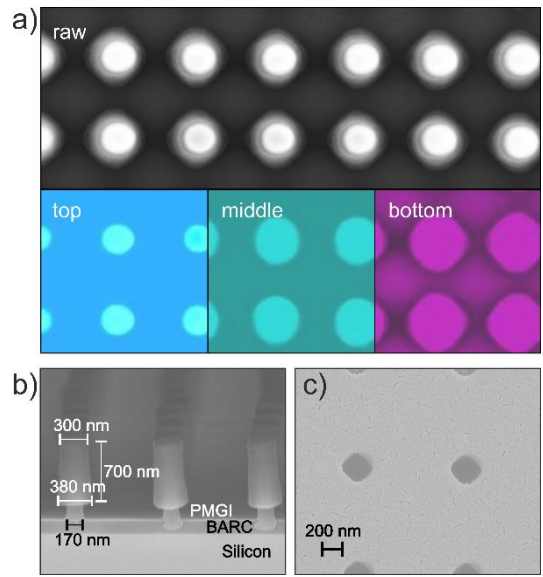

*Figure 4.* (a) Plan view SEM image of a dots array sample prepared with double DTL exposure method. The upper part of the image shows the raw SEM image. The different sections (top, middle and bottom part) of the photoresist pillars can be clearly distinguished in topographic contrast. Each part is highlighted with artificial filters to better show the change in size and shape from top towards the bottom of the tip. (b) Cross sectional SEM image of typical photoresist frustums after double DTL exposure and development; (c) plan view SEM of typical platinum pattern after deposition, lift-off process and thermal treatment.

The SEM image in Fig. 4a shows the plan view of a dot array in photoresist. The dots have frustum's shape (Fig. 4b), so the topographic contrast of SEM with secondary electrons signal can be used to highlight the different sections of the pillar with brightness decreasing from the top to the bottom. Artificial filters in plan-view SEM have been used to enhance the edge of top, middle, and bottom part of the photoresist frustums. From bottom to top, the size of the photoresist frustums decreases, and the shape gradually change from more square-like to more roundish. The laser light is absorbed by the thick photoresist layer as it propagates from the top to the bottom. Consequently, the upper part of the photoresist receives a higher dose if compared to the lower part of the photoresist. According to the simulation results, with a higher dose the frustums become smaller and more roundish.

**Pt pattern for gas-MacEtch**
A mask with linear grating 2 μm pitch and 0.5 duty cycle was used in this work for the preparation of the dots array pattern. Due to the Talbot effect, the dose distribution map has a period of 1 μm in the photoresist. The dots have diameter of approximately 300 nm after development (Fig. 4.b), and the diameter was further reduced to 200-250 nm after BARC removal by oxygen plasma. Thanks to the long exposure time, the size and shape of the photoresist pillars at top and bottom part are favorably similar (see details in Material and Methods) and differs only by 80 nm (Fig. 4.b). The platinum layer was deposited after BARC removal and the presence of undercut PMGI layer ensures a clean and neat lift-off process, leaving holes with sharp edges in the platinum layer, as shown in Fig. 4.c. The oxygen plasma treatment can be used to further reduce the dots size. We used an oxygen plasma process in reactive ion etching (RIE) and we estimated a consumption rate of 100 nm/min in the direction perpendicular to the photoresist frustums and 10 nm/min at the sidewalls. Since the photoresist layer has a thickness of about 700 nm after development (Fig. 4.b), and the thickness of Pt film for MacEtch is in the range of 10-12 nm, the RIE treatment can be tuned to further reduce the dots size without compromising the lift-off quality. We demonstrated successful lift-off for dots size in the Pt pattern of about 113 nm, which is 1/3 of the original dots size in the photoresist pattern (see details in Material and Methods) and 10 times smaller than the feature size in the photomask used for DTL.

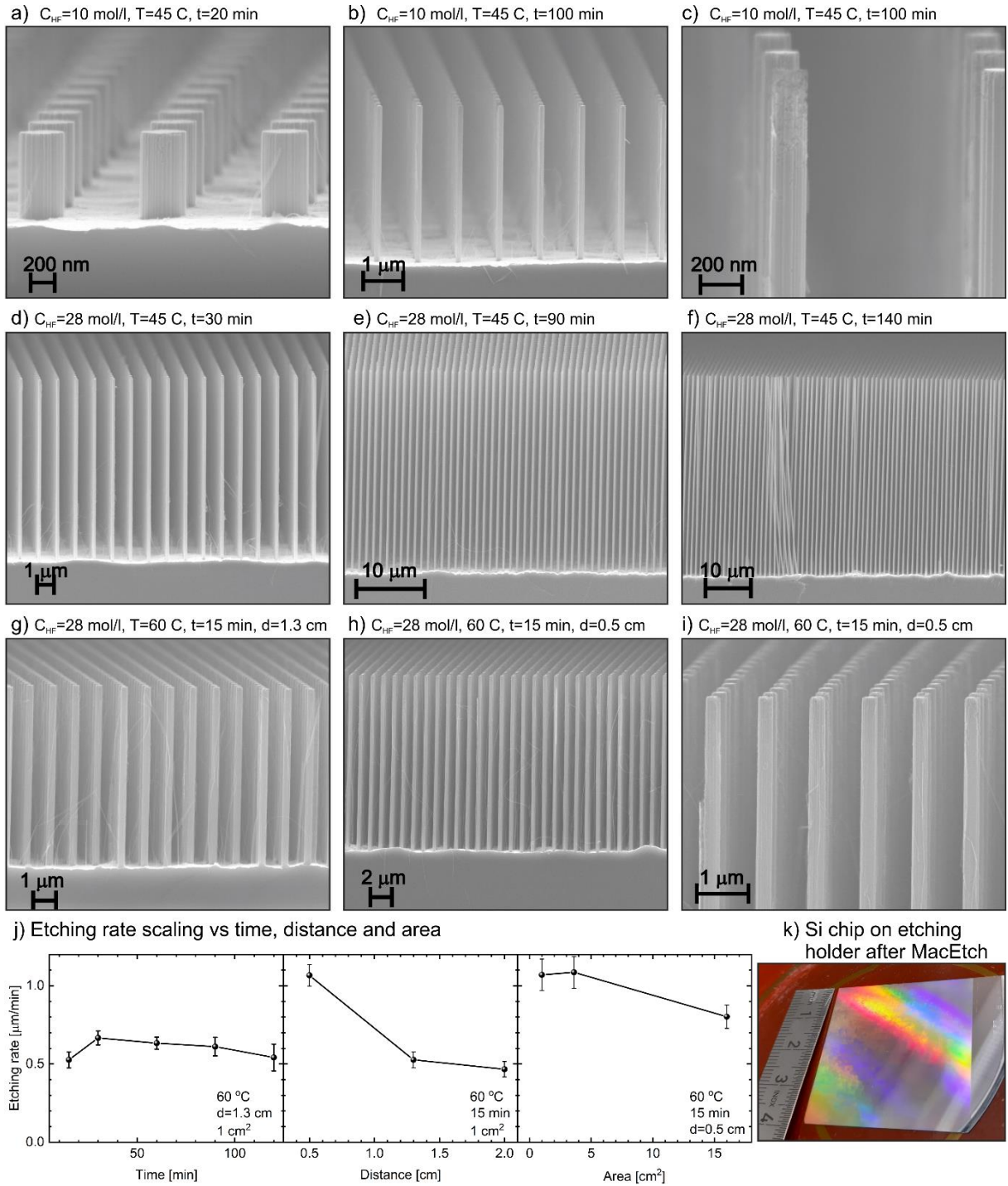

*Figure 5.* Cross sectional SEM images (a-i) of silicon nanopillars after gas-MacEtch at different etching rates, the starting Pt pattern have holes with size in the range of 200-300 nm. MacEtch with low HF content ($C_{HF}$=10 mol/l) in the liquid solution, the patterned Si chip of 1×1 cm² is at T=45 °C, the distance d between the Si chip and the liquid HF solution is d=1.3 cm: (a) nanopillar height h=730 nm, etching time of t=20 min; (b) h=3.9 µm, t=100 min. (c) high magnification of (b). MacEtch with high HF content ($C_{HF}$=28 mol/l), T=45 °C, d=1.3 cm: (d) h=10 µm, t=30 min; (e) h=29 µm, t=90 min; (f) h=43 µm, t=120 min. MacEtch with high HF content (28 mol/l), T= 60 °C, t=15 min, (g) d=1.3 cm, h=7.9 and (h) d=0.5 cm, h=16. (i) high magnification of (h). (j) Etching rate as a function of etching time, distance d and sample area, details are reported in Table 1. (k) image of patterned Si chip 4×4 cm² on the holder just after the gas-MacEtch process, light diffraction from periodic structures indicates the pillars are well ordered.

**Gas-MacEtch process**

The silicon nanopillars are created by MacEtch in gas-phase in a <100> Si substrate. The Pt patterned Si chip of 1×1 cm$^2$ is warmed up at a temperature T (45 and 60°C) by a back contact hot plate during the process and held at a distance d (0.5 – 2.0 cm) from the liquid solution of concentrated HF (5 – 28 mol/l). The HF evaporates from the liquid solution providing the etchant gas, while the oxygen is supplied from the air flowing inside the etching chamber. The MacEtch reaction occurs at the Pt layer (Eq. 2) with the injection of charge carriers in silicon, according to the literature silicon can be dissolved with two different reactions, the direct dissolution (Eq. 3), or via Si oxidation (Eq. 4), followed by the dissolution of the oxide (Eq. 5):

$$O_2 + 4H^+ + 4e^- \rightarrow 2H_2O \quad (2)$$

$$Si + 4h^+ + 4HF \rightarrow SiF_4 + 4H^+ \quad (3)$$

$$Si + 2H_2O + 4h^+ \rightarrow SiO_2 + 4H^+ \quad (4)$$

$$SiO_2 + 2HF_2^- + 2HF \rightarrow SiF_6^{2-} + 2H_2O \quad (5)$$

where e$^-$ indicates electrons and h$^+$ indicates holes injected in the Si substrate. The presence of air flow helps to release and disperse the etching by-products, acting as gas carrier, water is produced (Eq.2, 5) and released as vapor, while Si fluoride species (Eq. 3, 5) can be released as SiF$_4$ in gas phase[52].

In previous report[40], we realized the nano-fabrication of Si nanowires by gas-MacEtch in self-assembled Pt nanostructured patterns on <100> Si substrates. Here, we demonstrate that Si nanopillars with controlled shape by a lithographic pattern can be realized with the same process and on high resistivity <100> Si substrates (1-30 Ωcm). Very thin nanowires can be observed (see Fig. 5 a, b, d, g) because of the Pt pattern de-wetting. The thermal treatment of the Pt layer produced the platinum silicide, which stabilize the catalyst layer on top of the silicon substrate before the etching. The de-wetting temperature is very low (250 °C) to minimize the formation of Pt film cracks, which turn in very thin nanowires during the MacEtch. These nanowires are thinner than 2 nm and barely visible in the SEM images of long etching and heavily concentrated HF (Fig. 5 f), indicating they are almost completely consumed probably due to the Si oxidation reactions (Eq. 4-5). We demonstrate here an easy fabrication of Si nanopillars with a good control of the pattern transfer at different etching rate. The etching rate can be varied by changing the composition of the HF liquid solution (C$_{HF}$), the temperature of the Si chip (T) and the distance (d) between the Si chip and the surface of the liquid HF solution. Several etching conditions have been realized, details are reported in Table 1 (section Materials and Method).

The low etching rate of 40 nm/min has been realized by decreasing the HF content in the liquid solution (C$_{HF}$=10 mol/l), etching for 10 min produces nanopillars with height of 185 nm (Fig. S1 b), indicating an incubation time with a non-stable etching rate in this time frame. The high magnified SEM images of just formed Si nanopillars at low etching rate (Fig. 5 a, b) show the fidelity of the pattern transfer at the nanoscale. The roughness of the Si pillars is produced by the nanostructured contour of the holes in the Pt layer, which is visible in the SEM images in plan-view of the Pt pattern (Fig. 4 c), indicating a pattern transfer with nanometric resolution. The etching rate observed in this study are comparable to what previously reported for random nanowires pattern[40], indicating that the general trend is reproducible and not depending on the specific pattern structure. The etching rate can be further pushed down to 10 nm/min by decreasing the HF concentration at C$_{HF}$=5 mol/l. Since the gas-MacEtch reaction occurs in dry, the etched Si nanopillars are vertically standing and do not require any additional post-processing, such as critical point drying, which is necessary in case of wet MacEtch.

The distance of the Pt patterned Si chip from the surface of the liquid HF solution has a relevant impact on the etching rate, by reducing the distance to d=0.5 cm, the etching rate is doubled. The temperature has been increased to 60 °C to prevent the condensation of the liquid. The MacEtch reaction is also temperature dependent[44] with a slight increase of the etching rate from 45 to 60 °C. The difference in the HF concentration affects the quality of the nanopillars on long etching time. The injected charge carriers diffuse in the whole Si substrate and can reach the nanopillar surface too. Since the vapor is richer in water at lower HF concentration, the oxidation of the Si nanopillars is also favored and a relevant erosion at the top of the nanopillars is observed after long etching time (Fig. 5c). This phenomenon is present also in the case of high HF content, but it is less effective and results in more solid nanopillars even after long etching time (Fig. 5 f).

In summary, the overall quality of pattern transfer is maintained for all the conditions reported in Fig. 5 and Table 1, the etching rate can be tuned for the specific target height. High etching rate (0.3-0.5 µm/min) and very high aspect ratio (~200) can be achieved by increasing the HF content, high temperature and short distance are indicated for very fast processing (etching rate ~1 µm/min), while low etching rate can be targeted by lowering the HF content when a more precise control of the height is needed in the range of 0-3 µm (etching rate ~10nm/min for $C_{HF}$=5 mol/l). The processing time is in the range of 2 hr for aspect ratio up to 200.

**Challenges for nano-fabrication using DTL in conjunction with gas-MacEtch**
Although nanometric resolution pattern design capability in 2D, height controllability in the z-direction up to aspect ratio of 200, accessibility under ambient (fab-free) conditions, and relatively low processing cost are available in the fabrication method for Si nanopillars through DTL in conjunction with gas-MacEtch in <100> Si substrates, the method still has some limitations to its use in high-volume applications. The resolution and aspect ratio of the Si structures fabricated through the DTL in conjunction with gas-MacEtch are influenced by both techniques, since the DTL is related to 2D patterning and the gas-MacEtch enables the 2D patterns to extend to the z-direction. DTL patterning is available in our tool up to 8-inch wafer scale, but it is limited to periodic arrays[32,33]. In this work, we demonstrated the pattern transfer in a thin Pt film at nanometric resolution with positive resist and the help of a lift-off layer. The clean lift-off process is required for a high-fidelity pattern transfer into the Si substrate via MacEtch.

The obtained aspect ratio of the Si structures is restricted by the critical factors of the MacEtch process; the balance between the generated and injected hole carriers, the stability of the metal catalyst film, and the uniform supply of etchants strongly influence the stable Si etching in the vertical direction. Even though the DTL pattern was produced on a 4-inch wafer, we observed that the etching rate in gas-MacEtch is area dependent, with etching rate slowing down as a function of the size of patterned area. For example, we estimated a reduction of etching rate of 25 % by increasing the patterned area from 1 cm$^2$ to 16 cm$^2$. The etched depth is uniform on the whole etched pattern at distance of the at least 50 µm from the edge of the silicon chip[40]. This does not prevent to etch full 4-inch wafers in gas-MacEtch by increasing the etching time or the supply of etchants to consider the increase of the patterned area.

Since decreasing the distance between the liquid and the patterned Si substrate has the effect of speeding up the etching, the available concentration of HF in the vapor is depleted in the region of the sample and not enough to ensure the fast etching on larger samples. With increasing of aspect ratio, the reactants gases need to diffuse through the already etched structures to reach the catalyst at the bottom of the Si nanopillars, so the etching process becomes diffusion limited. We observed the formation of local defects in the catalyst pattern with etching depth unevenness after long etching time at very high speed. This depth unevenness usually develops in etching along non-<100> directions and deteriorate the vertical profile of the Si nanopillar arrays and the uniform movement of the whole catalyst pattern sinking into the Si substrate. Local unevenness deforms the catalyst pattern and initiate the spiralizing catalyst motion inside the Si substrate[55]. Therefore, in the high etching rate regime (T~60 °C, d~0.5 cm, high HF concentration) it is more difficult to maintain the etching perpendicular to the substrate. Some examples of pattern deformation are reported in the Supplementary Information, we observed vertical pillars with maximum length of 65 µm but the overall pattern is deformed by the onset of non-vertical etching. In previous work[40], we reported vertical etching depth in the range of 100 µm for very thin nanowires using blank de-wetting Pt pattern. For lithographically patterned structures, reaching aspect ratio higher than 200 in conditions of vertical etching is more challenging, indicating that the etching conditions need to be tuned for the specific pattern. The observed off-vertical etching direction after long etching in the high etching rate regime could be a consequence of the degradation of the HF solution as a function of time during the etching process. Therefore, the control of HF concentration in the liquid is critical for the reproducibility of the etching process as well as the cleaning conditions of the substrate before the Pt deposition, since residuals of the patterning process can prevent the adhesion[40] of Pt layer on the substrate with consequent instabilities or etching unevenness.

A refined instrument is necessary to maintain constant conditions of the HF vapor and to favor the diffusion of the reactants through the etched structures to ultimately maintain a uniform supply of the etchants at the catalyst surface with increasing the aspect ratio and the patterned area. Therefore, advanced gas-MacEtch equipment that increases the aspect ratio in the z-direction and the patterned area must be developed to utilize the DTL in conjunction with gas-MacEtch for high-value applications at full wafer scale. The advantage of combing gas-MacEtch with DTL patterning is a viable pathway to massively fabricate 3D protruding Si nanopillar arrays, with a faster nanopatterning with respect to scanning techniques[47], higher aspect ratio at the nanoscale with respect to RIE methods and no pattern collapse due to stiction with respect of wet-MacEtch.

## Conclusions

A straightforward method to scale up the nanofabrication of 3D protruding Si nanopillars arrays was demonstrated via DTL combined with z-directional etching depth-controllable MacEtch under gas-phase conditions. The approach of double exposure DTL, lift-off and final oxygen plasma etching allows patterning holes arrays in a thin Pt film with feature size (~100 nm) 10 times smaller than the lithography mask feature. The dose distribution map in DTL double exposure of linear grating masks was simulated by considering a Gaussian-like laser spectrum distribution. Due to the smearing effect, the second exposure allows to produce a pattern of pillars with rounded shapes and shrieked features's size with respect to the original photomask. The photoresist profile is controlled by tuning the exposure dose, the development time and a subsequent oxygen plasma treatment. The pattern transfer into silicon substrates is demonstrated with a nanoscale resolution. We showed that the etching rate in gas-MacEtch can be tuned by changing the HF content, the temperature, and the distance from the liquid HF solution. The etched Si nanopillars have vertical and solid profiles in high resistivity (1-30 Ωcm) N-type <100> wafers with etching rate up to 1 µm/min at 60 °C. It was observed that the Si porosity is increasing in the top part of the etched structures with increasing the etching time in conditions of low HF content. The distance of the Pt patterned Si chip from the surface of the liquid HF solution has a relevant impact on the etching rate, by reducing the distance to d=0.5 cm, the etching rate is doubled. Importantly, we showed that an array of vertical protruding Si structures with a width as low as 200 nm and a length of up to 43 µm can be obtained using the developed method without the stiction typically observed in wet-MacEtch process. With the used etching conditions of gas-MacEtch it is possible to achieve an aspect ratio in order of 200 in 2 hours. The advantages of gas-MacEtch are the relatively cheap processing, the low crystallographic damage of silicon substrates, the extreme aspect ratio at nanometric scale and hierarchical structuring capability in comparison to conventional plasma etching methods. There are significant needs of nanofabrication of Si nanopillars arrays in several relevant fields, such as photonic, phononic and biological metamaterials, batteries, X-ray optics, etc. Thus more research is necessary on scaling up the gas-MacEtch of arrays of Si nanostructures arrays on large area with a more efficient supply of reactants and their applications.

## Materials and methods

**Simulation**
The dose distribution map was simulated with Matlab following the method described below. A monochromatic plane wavefront with a wavelength of 377 nm passes through the quartz phase mask, where a periodic π phase shift is introduced to the wavefront. Consequently, an interference pattern appears downstream the phase mask, along the laser propagation direction. The wavefront $\psi(x, y, z)$ at a distance $z$ from the phase mask can be described with the Fresnel diffraction integral:

$$\psi(x, y, z) = \frac{e^{-ikz}}{i\lambda z} \iint \psi(x, y, 0) e^{\frac{-ik}{2z}[(x-x_0)^2 + (y-y_0)^2]} dx_0 dy_0 \quad (6)$$

where $\psi(x, y, 0)$ is the form of the wavefront right after it passes through the phase mask, $x$ and $y$ are the coordinates in the plane that is parallel to the phase mask plane, $x_0$ and $y_0$ are the corresponding $x$ and $y$ values in the plane right after the phase mask plane, and $k$ is the wave number ($2\pi/\lambda$). The Fresnel propagator is defined as:

$$h(x, y, z) = \frac{e^{-ikz}}{i\lambda z} e^{\frac{-ik}{2z}(x^2 + y^2)} \quad (7)$$

The integral is then expressed as a convolution:

$$\psi(x, y, z) = \psi(x, y, 0) * h(x, y, z) \quad (8)$$

According to the convolution theorem, Eq. 8 can be transformed into Eq. 9:

$$\psi(x, y, z) = Ft^{-1}\{Ft\{\psi(x, y, z)\} \cdot Ft\{h(x, y, z)\}\} \quad (9)$$

The intensity map $I(x, y, z)$ at position z is obtained by taking the absolute value of the square of the complex wave function:

$$I(x, y, z) = |\psi(x, y, z)|^2 \quad (10)$$

The DTL process is completed by scanning the wafer along the laser propagation direction for a scanning range of $r = n \times z_T$. The total dose $D_m$ collected by the wafer over the scanning range $r$ is defined by scanning time $t_s$ and is calculated with the following formula:

$$D_m = \int_0^{t_s} I(x, y, z_t) dt \quad (11)$$

Considering that instead of monochromatic, the laser has a Gaussian like spectral distribution, we then apply the Gaussian filter $g(x, y)$ to the dose distribution map, the new dose distribution map $D_G$ is then calculated:

$$D_G = D_m * g(x, y) \quad (12)$$

Then the final dose distribution map $D$ after the double DTL exposure process is calculated and plotted with the following expression:

$$D = D_G + D_G^T \quad (13)$$

where $D_G^T$ is the transpose of the dose distribution map $D_G$ after the first DTL exposure process.

**Nano-fabrication**
The 4 inch N-type boron doped <100> wafers (1-30 Ωcm) are pre-cleaned with $O_2$ plasma in a TePla wafer asher for 10 min. The gas flow is set at 600 sccm, and the power is set at 600 W. A layer of AZ BARLi-II BARC is spin coated at a speed of 4000 rpm and baked at a temperature of 180 °C for 1 min, resulting in a thickness of approximately 200 nm, in order to have a low reflectivity for i-line light sources. MicroChem PMGI-5S resist is spin coated at a speed of 4000 rpm, resulting in a thickness of around 150 nm; the soft baking temperature in the range of 145 to 160 °C is selected to ensure sufficiently high undercut for the following lift-off process. SUMIRESIST PFI-88A7 positive photoresist is spin coated at a speed of 4000 rpm and baked at a temperature of 90 °C for 1 min, and the thickness of the photoresist layer is around 700 nm. The DTL exposure is performed with a Eulitha PhableR200 system. A polarized laser light with a wavelength of 377 nm is used for the exposure process. Linear phase grating mask with a period of 2 μm introduce a periodic π phase shift to the laser wavefront, which produces an interference pattern. The laser intensity is set at 0.5 mW/cm$^2$. The wafer scanning range is 105 μm, which corresponds to three times the Talbot distance, for the purpose of reducing the beating effect[33]. The total exposure time for one DTL process is 200 s, so that the energy density is 0.1 mJ/cm$^2$. The wafer is then rotated by 90 deg, and a second exposure is performed with the same set of parameters as the first exposure. After a 1 min post exposure baking at 110 °C, the wafer is developed in Megaposit MF-24A developer for 1 min. The PMGI layer is developed together with the photoresist with a certain level of undercut.

To generate a dot array that has a sparse distribution and the smallest possible dot diameter, a relatively long exposure time is set for the process. To improve the exposure quality and have better control over the exposure profile, a series of exposure experiments were carried out with a linear photomask that has a period of 2.4 μm, so the period in the exposed photoresist is 1.2 μm. The purpose of this test is to learn how the exposure profile changes with the exposure time, so the patterns were not prepared for lift-off (no PMGI layer between the BARC and PR layers). The widths of the trenches are measured at both top and bottom of the photoresist layer on the exposed sample (Fig. 6.a). Figure 6.b reports the data of the measured profile as a function of the exposure time at different development time. It is noticeable that the development time has a minor effect on the profile, while the exposure time has a more obvious effect on the final profile. As a general trend, the trench gets wider at both top and bottom of the photoresist, but the width difference between top and bottom becomes smaller as the exposure time increases.

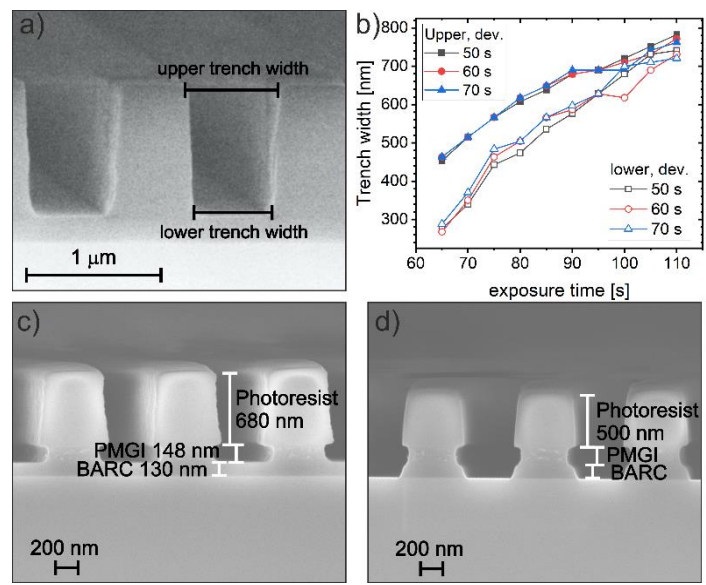

*Figure 6.* The laser is absorbed as it propagates towards the bottom of the photoresist layer, so the actual accepted dose in the upper part of the photoresist is relatively higher than the bottom part. The difference is higher when exposure time is shorter and becomes smaller when the exposure time is longer. (a) Cross sectional SEM image of the photoresist layer (grating with pitch 1.2 μm, single exposure). (b) The widths of the trenches in the photoresist layer measured at upper and lower part as a function of different exposure time and development time. Cross sectional SEM image of the stuck photoresist/PMGI/BARC (grating with pitch 1 μm, single exposure of 220 s) (c) after the development time of 60 s and (d) after the BARC removal by oxygen plasma.

The BARC layer is etched with $O_2$ plasma in an Oxford RIE etcher, so that the underneath Si substrate is exposed. This treatment is also cleaning the eventual residual of photoresist within the trenches before the subsequent metal deposition. We deposited the Pt catalyst layer by electron beam assisted evaporation. The film thickness is in the range of 10-12 nm. After the deposition, the lift-off is carried out in acetone at room temperature in an ultrasounic bath. After lift-off the Pt film is annealed at 250 °C for 30 min in air in order to stabilize the Pt layer for the subsequent MacEtch process with minimal de-wetting (small cracks in Pt film, see Fig. 4 c) and the formation of a platinum silicide layer, according to previous reports [39,40].

The $O_2$ plasma RIE treatment can be used to further reduce the photoresist frustums size to obtain smaller holes in the Pt film. Figure 6 reports some examples of shrinking by increasing the RIE etching time. The effective shrinking is measured directly in the Pt pattern after lift-off since this is the most critical step. The RIE etching conditions are: RF power of 100 W, 10 sccm $O_2$, chamber pressure of 20 mTorr. We estimated an etching rate of 100 nm/min in the direction perpendicular to the substrate and 10-17 nm/min from the sidewalls of the photoresist pillars. The etching rates depend on the RIE tools and machine conditioning, so they are only indicative of the required process tuning. The remaining photoresist is sufficient to obtain a successful lift-off of the Pt holes up to 113 nm, which is about 1/3 of the original size of the photoresist pillars.

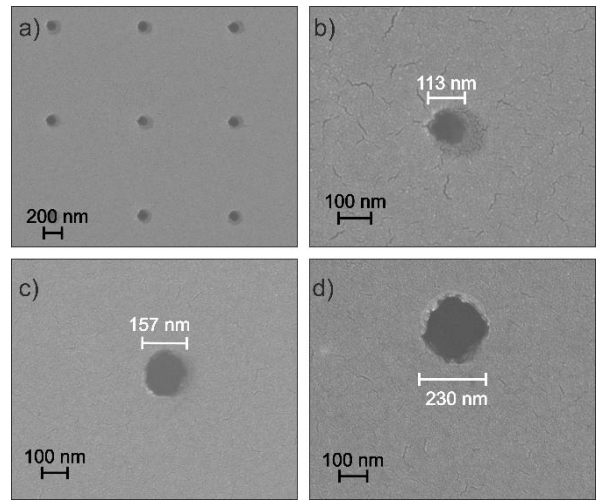

*Figure 7*. SEM in plan-view of the Pt patterned film after lift-off for different etching time of $O_2$ plasma in RIE: (a) 6 min; (b) 6 min, high magnification of (a); (c) 4 min; (d) 2 min.

The gas-phase MacEtch is performed in a vapor HF tool by Idonus Sarl according to previous report[40]. A patterned chip (typically of 1×1 $cm^2$) is cleaved out the patterned wafer. The liquid container of the vapor HF tool is filled up to the top with 200 ml of HF solution containing HF at 50 wt% in volume and deionized water (18 MΩ cm). The patterned chip is kept at a distance d in the range of 0.5 – 2 cm from the liquid solution and the air flowing inside the chamber provides the oxygen for the MacEtch reaction at the platinum catalyst layer. The pattern chip is warmed during the gas-MacEtch at a temperature of 45 and 60 °C. Since the HF liquid concentration can be degraded by repetitive warming cycles, especially in short distance conditions, the HF liquid solution was freshly prepared for each experiment.

Table 1 summarizes the etching conditions ($C_{HF}$, T, t, d), the measured nanopillars height and aspect ratio (the nanopillars width is about 200 nm), the etching rate observed in this work according to the data of Fig.5, Fig. S1 and Fig. S2.

*Table 1.* Etching rate of gas-MacEtch for different conditions of temperature of the Si chip (T), distance (d) between the Si chip and the surface of the liquid HF solution, etching time (t), HF concentration in the liquid solution. The measured Si nanopillars height (h) and the aspect ratio (AR) are also indicated.

| h [μm] | AR | Etching time [t] | Etching rate [μm/min] | T [°C] | d [cm] | $C_{HF}$ [mol/l] | Figure reference |
|---|---|---|---|---|---|---|---|
| 0.28 | 1 | 30 | 0.01 | 45 | 1.4 | 5 | Fig.S1.a |
| 0.19 | 1 | 10 | 0.02 | 45 | 1.3 | 10 | Fig. S1.b |
| 0.73 | 3 | 20 | 0.036 | 45 | 1.3 | 10 | Fig.5.d |
| 1.1 | 4.4 | 30 | 0.037 | 45 | 1.3 | 10 | Fig.S1.c |
| 3.9 | 20 | 100 | 0.039 | 45 | 1.3 | 10 | Fig.5.e, f |
| 10.0 | 50 | 30 | 0.33 | 45 | 1.3 | 28 | Fig.5.g |
| 29.7 | 149 | 90 | 0.33 | 45 | 1.3 | 28 | Fig.5.h |
| 43.0 | 215 | 140 | 0.31 | 45 | 1.3 | 28 | Fig.5.i |
| 7.9 | 40 | 15 | 0.53 | 60 | 1.3 | 28 | Fig.5.j |
| 16 | 80 | 15 | 1.07 | 60 | 0.5 | 28 | Fig.5.k, l |
| 38 | 190 | 60 | 0.63 | 60 | 1.3 | 28 | Fig.S2.a |
| 65 | 325 | 120 | 0.54 | 60 | 1.3 | 28 | Fig.S2.b |

The height of Si nanopillars has been measured by SEM in cross section. Since the etching rate is faster at the edge of the patterned Si chip[40], the images have been acquired in the center of the Si chip.

**SEM Imaging**
The SEM images were taken using the In-Lens detector of a Zeiss Supra VP55, the cross sections have been realized by cleaving the silicon substrate and the SEM images have been acquired at 5 deg tilt from the perpendicular to the cleaved surface. Therefore, the SEM cross-sectional images do not suffer distortions or artifacts typically observed in ion milling. However, the cleavage can also affect the very high aspect ratio Si nanopillars (Fig. 5 i) showing some local distortions of the nanopillars very close to the sample cleavage, that has been not observed in other locations. The high magnification images in plan-view in Figure 7 have been acquired at 5 keV with in-lens detector and working distance of 3 mm.

## Acknowledgements

We would like to thank LMN-PSI for clean room facilities; D. Marty (PSI), V. A. Guzenko (PSI), C. Wild (PSI) and Gordan Mikuljan (PSI) for technical support.

We acknowledge the support from: GratXray AG; European Research Council (310005); Schweizerischer Nationalfonds zur Förderung der Wissenschaftlichen Forschung (159263, 206021_177036, 206021_189662, CRSII2_154472, CRSII5_183568); lottery fund SwissLOS of the Kanton of Aargau; NanoArgovia Grant 13.01 ''NANOCREATE'' (Swiss Nanoscience Institute).

# Supplementary Information

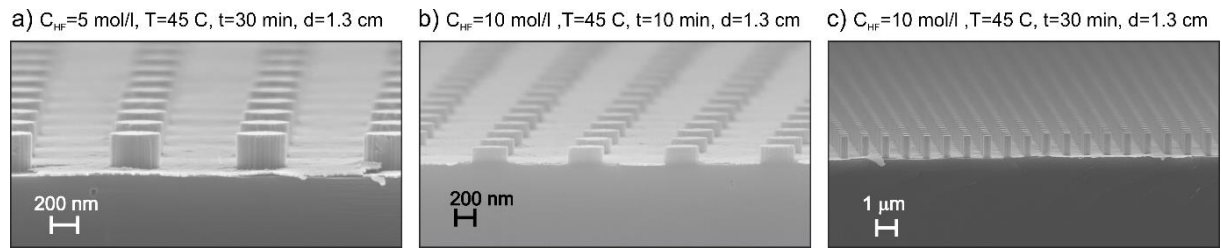

*Figure S1.* Cross sectional SEM images after gas-MacEtch at T=45 °C, d=1.3 cm with low HF concentration in the liquid solution: (a) $C_{HF}$= 5 mol/l, h=280 nm, t=30 min ; (b) $C_{HF}$=10 mol/l, h=185 nm, t=10 min  (c) $C_{HF}$=10 mol/l, h=1.1 μm, t=30 min.

In low HF content conditions ($C_{HF}$=5, 10 mol/l), the etching rate can be slowed down to 10-40 nm/min, allowing a very good control of the pattern transfer with nanoscale resolution (Fig. S1).

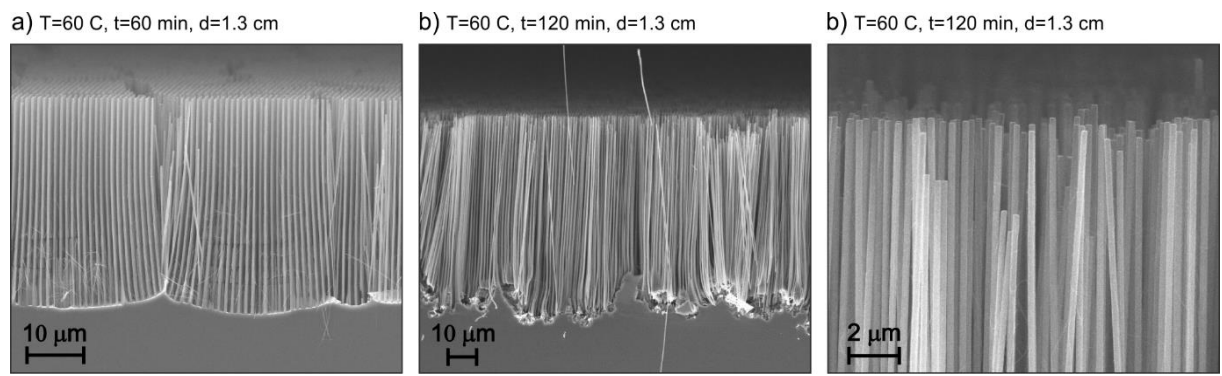

*Figure S2.* Cross sectional SEM images after gas-MacEtch with high HF content ($C_{HF}$=28 mol/l), T=60 °C, d=1.3 cm: (a) h=38 μm, t=60 min; (b) h=65 μm, t=120 min; (c) high magnification of (b).

The high etching rate conditions ($C_{HF}$=28 mol/l, T=60 °C) cause the unevenness of the Pt pattern (Fig.S2 a), which degenerate in etching along non-<100> directions after long etching process (Fig. S2.b). Despite the out of vertical etching, the Si nanopillars are still well separated with low porous top profile (Fig. S2.c).